\documentstyle[12pt]{article}
\newcommand{\be}{\begin{equation}}
\newcommand{\ee}{\end{equation}}
\newcommand{\bea}{\begin{eqnarray}}
\newcommand{\eea}{\end{eqnarray}}
\newcommand{\ba}{\begin{array}}
\newcommand{\ea}{\end{array}}
\newcommand{\bmat}{\left(\ba}
\newcommand{\emat}{\ea\right)}

\newcommand{\norsl}{\normalsize\sl}

\begin{document}
\begin{titlepage}
\title{
Super 5-branes in $D=10\ N=1$ \\
Super Yang-Mills Theory
}

\author{ \\\normalsize
  Yoshinori GOTOH\footnote{E-mail:\ gotoh@phys.h.kyoto-u.ac.jp} 
 and  Ikuo OKOUCHI\footnote{E-mail:\ dai@phys.h.kyoto-u.ac.jp} \\
\norsl  Graduate School of Human and \\
\norsl  Environmental Studies, Kyoto University\\
\norsl  Kyoto 606-01, JAPAN\\
\\
}

\date{}

\maketitle

\begin{abstract}
{\normalsize
In $D=10$ $N=1$ super Yang-Mills theory 
we give the background breaking a half of supersymmetry.
In the background there is a six-dimensional 
object so called $D=10\ $ 5-brane.
}
\end{abstract}

\begin{picture}(5,2)(-310,-500)
\put(35,-105){KUCP-0116}
\put(35,-120){hep-th/9804073}
\put(35,-135){Apr. 1998}
\end{picture}

\vspace{2cm}
 
\thispagestyle{empty}
\end{titlepage}
\setcounter{page}{1}
\baselineskip 20pt

%
%
\vspace{0.3cm}
\leftline{\large\bf 1. Introduction}
\vspace{0.3cm}
%

The possibility of partial supersymmetry breaking was discussed by 
Hughes and Polchinski \cite{HP1}. They pointed out the way how 
to evade the existing no-go theorem of partial supersymmetry
breaking based on the supersymmetry current algebra. They 
constructed a supersymmetric Nielsen-Olesen vortex \cite{NO} in a $D=4\ N=2$
supersymmetric theory in which half of the supersymmetry is 
spontaneously broken. They also showed that the effective action for
the vortex is the Green-Schwarz covariant action \cite{GS}. Moreover 
Hughes, Liu and Polchinski \cite{HLP} gave a four-dimensional 
supermembrane solution\footnote{We would now call it 3-brane.}
of the six-dimensional Abelian $N=1$ 
supersymmetric gauge theory. They showed that the effective action 
for the membrane is a generalization of the Green-Schwarz covariant
action.

We give the background breaking a half of supersymmetry
in $D=10$ $N=1$ super Yang-Mills theory.
In the background there is a six-dimensional object so called $D=10$
5-brane. We also give the effective action 
for the 5-brane which is a generalization of the Green-Schwarz covariant
action as same as the case of \cite{HLP}.

In sect.2 we first introduce $\Gamma$-matrices in $D=10$ Minkowski 
space that will be important in the later sections. In sect.3
we consider dimensional reductions and review that $D=4$ $N=4$ 
super Yang-Mills Lagrangian can be derived
from $D=10$ $N=1$ Lagrangian by trivial dimensional reduction \cite{BSS}. 
In sect.4 we give a topological solution of 
$D=10$ $N=1$ super Yang-Mills theory which denotes a $D=10\ $ 5-brane.
This background effectively realizes $D=4$ $N=2$ 
super-Poincar{\'e} symmetry. Thus it has only 
half of the supersymmetry as compared with the case of trivial 
dimensional reduction. Finally we give the effective action 
for the 5-brane\footnote{P-brane action was shown in \cite{AETW}\cite{BSTT}}.

%
\vspace{0.3cm}
\leftline{\large\bf 2.  $\Gamma$-matrices in $D=10$ Minkowski space}
\vspace{0.3cm}
%

We introduce a particular representation of the $D=10$
$\Gamma$-matrices\footnote{We use the representation of the $D=10$
$\Gamma$-matrices of \cite{SOH}.}:
\be
\{\Gamma_M,\Gamma_N\}=2\eta_{MN}
\label{eqn:g1}
\ee
where $\eta_{MN}=(1,-1,...-1)$. We choose $\Gamma_M (M=0,..3,5,..10)$ as
\bea
&& \Gamma_\mu =\gamma_\mu \bigotimes \mbox{\Large 1} \ \ \ \ \ \ \ \ \
\ \ for\ \mu = 0,...3
\nonumber\\
&& \Gamma_{4+m}=\gamma_5 \bigotimes \tilde{\Gamma}_m \ \ \ \ \ for\ m=1,...6
\label{eqn:g11}
\eea
where $\gamma_\mu , \gamma_5$ the $4 \times 4\ \Gamma$ matrices in $D=4$ 
Minkowski space with $\eta_{\mu \nu}=(1,-1,-1,-1)$ and $\tilde{\Gamma}_m$ 
the $8 \times 8\ \Gamma$ matrices in $D=6$ Euclidean space with $\eta_{mn}=
\delta_{mn}$
\bea
&&\gamma_\mu =
\bmat{cc}
0 &  (\sigma_\mu)_{\alpha \dot{\alpha}}\\
(\bar{\sigma}_\mu)^{\dot{\alpha}\alpha} & 0\\
\emat
\ \ \ \ \ 
\gamma_5 =
\bmat{cc}
-i&0\\
0&i\\
\emat
\nonumber\\ 
&&\tilde{\Gamma}_m =
\bmat{cc}
0 &  (\tilde{\sigma}_m)_{ij}\\
(\tilde{\sigma}^{-1}_m)^{ij} & 0\\
\emat 
\label{eqn:g2}
\eea
where $i,j=1,...4$, Pauli-matrices \boldmath$\sigma$\unboldmath=
$(\sigma_1,\sigma_2,\sigma_3),\  
\sigma_\mu =(1,$\boldmath$\sigma$\unboldmath$), 
\bar{\sigma}_\mu =(1,-$\boldmath$\sigma$\unboldmath$)$ and 
\bea
&&
\tilde{\sigma}_1 =\bmat{cc}
0 & - \sigma_3 \\
\sigma_3& 0\\
\emat
\ \ \ \ \ 
\tilde{\sigma}_2 =\bmat{cc}
0 & i \\
-i& 0\\
\emat
\ \ \ \ \ \ 
\tilde{\sigma}_3 =\bmat{cc}
0 & \sigma_1 \\
-\sigma_1& 0\\
\emat\nonumber\\
&&
\tilde{\sigma}_4 =\bmat{cc}
0 & - \sigma_2 \\
-\sigma_2& 0\\
\emat
\ \ \ 
\tilde{\sigma}_5 =\bmat{cc}
- \sigma_2&0 \\
0&\sigma_2\\
\emat
\ \ \ 
\tilde{\sigma}_6 =\bmat{cc}
i\sigma_2&0 \\
0&i\sigma_2\\
\emat.
\label{eqn:g3}
\eea
These $\tilde{\sigma}_m$ satisfy the following relations:
\bea
&&(\tilde{\sigma}_m)_{ij}=-(\tilde{\sigma}_m)_{ji},\ \ 
(\tilde{\sigma}^{-1}_m)^{ij}=(\tilde{\sigma}^{-1}_m)^{ji}\nonumber\\
&&(\tilde{\sigma}_m)_{ij}^{\ast}\equiv -(\tilde{\sigma}^{-1}_m)^{ij}
=\frac12\epsilon^{ijkl}(\tilde{\sigma}_m)_{kl}\nonumber\\
&&(\tilde{\sigma}_m)_{ij}(\tilde{\sigma}^{-1}_n)^{ji}=4\delta_{mn},\ \ 
(\tilde{\sigma}_m)_{ij}(\tilde{\sigma}^{-1}_m)^{kl}
=-2(\delta_i^k\delta_j^l-\delta_i^l\delta_j^k)\label{eqn:g4}\\
&&(\tilde{\sigma}_m)_{ij}(\tilde{\sigma}_m)_{kl}=2\epsilon_{ijkl},\ \ 
(\tilde{\sigma}^{-1}_m)^{ij}(\tilde{\sigma}^{-1}_m)^{kl}=2\epsilon^{ijkl}
\nonumber.
\eea
$\Sigma$-matrices are defined by
\bea
&&\Sigma_{MN}\equiv\frac{i}{2}[\Gamma_M ,\Gamma_N]
\nonumber\\
&&\Sigma_{\mu\nu}=\frac{i}{2}[\Gamma_\mu ,\Gamma_\nu]
=\frac{i}{2}[\gamma_\mu ,\gamma_\nu]\bigotimes \mbox{\Large
  1}\label{eqn:g5}\\ 
&&\Sigma_{mn}=\frac{i}{2}[\Gamma_m ,\Gamma_n]
=-\frac{i}{2}\mbox{\Large 1}\bigotimes [\tilde{\Gamma}_m ,\tilde{\Gamma}_n].
\nonumber
\eea

%
\vspace{0.3cm}
\leftline{\large\bf 3.  Trivial dimensional reduction}
\vspace{0.3cm}
%

Consider the $N=1$ super Yang-Mills Lagrangian in $D=10$ 
Minkowski space
for a gauge field $A_M\ (M=0,..3,5,..10)$ and a Majorana Weyl
spinor $\lambda$ in the adjoint representation of 
the gauge group\footnote{Supersymmetric matter contents in higher 
dimensions are given in \cite{STR}.}:
\be 
{\cal L}=-\frac{1}{4}F_{MN}^a F^{aMN} + \frac{i}{2}\bar{\lambda}^a
\Gamma^M D_M\lambda^a
\label{eqn:t1}
\ee
where the $D=10$ Majorana Weyl spinor $\lambda$ can be written by 
$D=4$ Weyl spinors as 
\be
\lambda=(\lambda_{\alpha i},0,0,\bar{\lambda}^{\dot{\alpha}i})^T
\ \ \ \ for\  \alpha, \dot{\alpha}=1,2\ and\ i=1,...4\ . 
\label{eqn:t2}
\ee
This Lagrangian is invariant under supersymmetry transformations:
\bea
\delta A_M^a &=& i\bar{\xi} \Gamma_M \lambda^a \nonumber\\
\delta \lambda^a &=& -\frac{i}{2}\Sigma^{MN}\xi F^a_{MN}
\label{eqn:t3}
\eea
where $\xi$ is a $D=10$ Majorana Weyl spinor supersymmetry parameter 
and can be written by $D=4$ Weyl spinors 
$(\xi_{\alpha i},0,0,\bar{\xi}^{\dot{\alpha}i})^T$.

To obtain a $D=4$ theory from this, we perform a trivial dimensional
reduction:
\be
\partial_{4+m}=0 \ \ \ \ \ for\ m=1,...6\ .
\label{eqn:t4}
\ee
It gives us 
\bea
&&F_{\mu \nu}=\partial_\mu A_\nu - \partial_\nu A_\mu - ig[A_\mu,A_\nu]
\nonumber\\
&&F_{\mu m}=\partial_\mu \phi_m - ig[A_\mu,\phi_m] \equiv D_\mu \phi_m\\
&&F_{m n}=- ig[\phi_m,\phi_n] \nonumber
\label{eqn:t5}
\eea
where $\phi_m \equiv A_m$. The Lagrangian (\ref{eqn:t1}) now depends
only on four coordinates $x^\mu$ and reads
\bea
-\frac{1}{4}F_{MN}^a F^{aMN}&=&-\frac{1}{4}F_{\mu \nu}^a F^{a\mu \nu}
+\frac{1}{2}D_\mu \phi_m^a D^\mu \phi_m^a 
+\frac{1}{4}g^2[\phi_m,\phi_n]^a[\phi_m,\phi_n]^a\nonumber\\
&=&-\frac{1}{4}F_{\mu \nu}^a F^{a\mu \nu}
+\frac{1}{2}D_\mu \phi_{ij}^a D^\mu \phi^{aij} 
+\frac{1}{4}g^2[\phi_{ij},\phi_{kl}]^a[\phi^{ij},\phi^{kl}]^a
\label{eqn:t6}\\
\frac{i}{2}\bar{\lambda}^a\Gamma^M D_M\lambda^a &=&
\frac{i}{2}\bar{\lambda}^a\Gamma^\mu D_\mu\lambda^a + 
\frac{1}{2}\bar{\lambda}^a\Gamma^m [\phi_m,\lambda]^a \nonumber\\
&=&\frac{i}{2}(\lambda^a_i\sigma^\mu D_\mu\bar{\lambda}^{ai}
+\bar{\lambda}^{ai}\bar{\sigma}^\mu D_\mu \lambda^a_i)
+gf^{abc}(\phi^{bij} \lambda_i^a \lambda_j^c + \phi^{b}_{ij}
\bar{\lambda}^{ai}\bar{\lambda}^{cj})
\label{eqn:t61}
\eea
where $\phi_{ij}$ and $\phi^{ij}$ are real anti-symmetric tensors defined by 
$\phi_{ij}\equiv -\frac{1}{2}(\tilde{\sigma}^m)_{ij}\phi_m$, 
$\phi^{ij}\equiv \frac{1}{2}(\tilde{\sigma}^{-1\ m})^{ij}\phi_m$ and 
$\phi_m = \frac{1}{2} (\tilde{\sigma}^{-1}_m)^{ij}\phi_{ij}
 = -\frac{1}{2}(\tilde{\sigma}_m)_{ij}\phi^{ij}$,
 and we have used the relations (\ref{eqn:g4}).
The supersymmetry transformations (\ref{eqn:t3}) become
\bea
\delta A_\mu^a &=& i\xi \Gamma_\mu \lambda^a 
= i(\xi_i\sigma_\mu\bar{\lambda}^{ai}+\bar{\xi}^i\bar{\sigma}_\mu\lambda^a_i)
\label{eqn:t7}\\
\delta\phi_m^a&=&i\xi \Gamma_m \lambda^a \label{eqn:t8}\\
\delta\phi_{ij}^a &=&\xi_i\lambda^a_j - \xi_j\lambda^a_i +
\epsilon_{ijkl}\bar{\xi}^k\bar{\lambda}^{al}\nonumber\\
\delta \lambda^a &=& -\frac{i}{2}\Sigma^{MN}\xi F^a_{MN}\label{eqn:t9}\\
\delta \lambda^a_{\alpha  i}&=&-\frac{i}{2}(\sigma^{\mu\nu}\xi_i)
_\alpha F^a_{\mu\nu}-2iD_\mu\phi_{ij}^a(\sigma^\mu\bar{\xi}^j)
_\alpha-2ig[\phi_{ij},\phi^{jk}]^a\xi_{\alpha k}\nonumber\\
\delta \bar{\lambda}^{a\dot{\alpha}i}&=&-\frac{i}{2}(\bar{\sigma}
^{\mu\nu} \bar{\xi}^i)^{\dot{\alpha}}F^a_{\mu\nu}
-2iD_\mu\phi^{aij}(\bar{\sigma}^\mu\xi_j)^{\dot{\alpha}}
-2ig[\phi^{ij},\phi_{jk}]^a\bar{\xi}^{\dot{\alpha} k}.\nonumber
\eea
The background (\ref{eqn:t4}) has the global $SO(1,3) \times SO(6) \sim 
SO(1,3) \times SU(4)_R$ symmetry. Actually 
these Lagrangian and transformations are known as the $SU(4)$ 
covariant $D=4$ $N=4$ super Yang-Mills theory.

%
\vspace{0.3cm}
\leftline{\large\bf 4.  Partial supersymmetry breaking and super 5-branes}
\vspace{0.3cm}
%

In this section we show that the super Yang-Mills Lagrangian in $D=10$ 
Minkowski space have topological solutions which break half of 
the supersymmetries.

We choose the following topological solutions:
\bea
&&A_\mu^a=0\quad\quad\quad\quad\quad\quad\quad\quad\quad\quad\ 
for\ \mu = 0,...3 \ \ \ (M=0,...3)\label{eqn:d1}\\
&&A_9^a,\ A_{10}^a=0\label{eqn:d11}\\
&&F^a_{\acute{m}\acute{n}}=\frac12
\epsilon_{\acute{m}\acute{n}\acute{k}\acute{l}}F^a_{\acute{k}\acute{l}}
\equiv \tilde{F}^a_{\acute{m}\acute{n}}\quad\quad\quad
for\ \acute{m},\acute{n},\acute{k},\acute{l}
=1,...4 \ \ \ (M=5,...8)\label{eqn:d12}\\
&&\lambda^a=0\label{eqn:d13}
\eea
where $F^a_{\acute{m}\acute{n}}$ depend only on four coordinates
$x^{\acute{m}}$.
 When we choose $SU(2)$ gauge symmetry, the topological background has
the global $SO(1,5) \times SO(3)_{diag}$ symmetry. Thus these solutions
denote a super 5-brane. 
 
To see partial supersymmetry breaking clearly we
assume that nothing depends on 2 dimensions:
\be
\partial_{9}=\partial_{10}=0,
\label{eqn:d2}
\ee
this background has the global 
$SO(1,3) \times SO(3)_{diag}\times SO(2)\sim SO(1,3) \times SU(2)_R\times 
U(1)_R$ symmetry thus $D=4$ $N=2$ supersymmetry.
We will expect that only half of the supersymmetries 
are unbroken in this background (\ref{eqn:d1})$\sim$(\ref{eqn:d2}) 
as compared with the case of trivial reduction of six dimensions 
(\ref{eqn:t4}). 
Actually we will show half of the supersymmetries are unbroken in the
topological background while the other half are broken. 

Before proceeding this for preparation 
we will investigate $\Sigma$-matrices (\ref{eqn:g5}).  
We define $\tilde{\Sigma}_{\acute{m}\acute{n}}$ as
\be
[\tilde{\Gamma}_{\acute{m}},\tilde{\Gamma}_{\acute{n}}]
\equiv 2i
\bmat{cccc}
\tilde{\Sigma}^1_{\acute{m}\acute{n}}&&&\\
&\tilde{\Sigma}^2_{\acute{m}\acute{n}}&&\\
&&\tilde{\Sigma}^3_{\acute{m}\acute{n}}&\\
&&&\tilde{\Sigma}^4_{\acute{m}\acute{n}}\\
\emat\\
\label{eqn:d8}
\ee
and find the following relation:
\bea
\frac12\epsilon_{\acute{k}\acute{l}\acute{m}\acute{n}}
[\tilde{\Gamma}_{\acute{m}},\tilde{\Gamma}_{\acute{n}}]&=&
\frac12\epsilon_{\acute{k}\acute{l}\acute{m}\acute{n}}\ 2i
\bmat{cccc}
\tilde{\Sigma}^1_{\acute{m}\acute{n}}&&&\\
&\tilde{\Sigma}^2_{\acute{m}\acute{n}}&&\\
&&\tilde{\Sigma}^3_{\acute{m}\acute{n}}&\\
&&&\tilde{\Sigma}^4_{\acute{m}\acute{n}}\\
\emat\nonumber\\
&=&2i
\bmat{cccc}
-\tilde{\Sigma}^1_{\acute{m}\acute{n}}&&&\\
&\tilde{\Sigma}^2_{\acute{m}\acute{n}}&&\\
&&-\tilde{\Sigma}^3_{\acute{m}\acute{n}}&\\
&&&\tilde{\Sigma}^4_{\acute{m}\acute{n}}\\
\emat.
\label{eqn:d9}
\eea
In the last step we have used the representations of 
(\ref{eqn:g2}) and (\ref{eqn:g3}).

Now we will see the supersymmetry transformations (\ref{eqn:t3}) in
the topological background. The supersymmetry variation of bosonic
fields is zero. The supersymmetry variation of fermion field is 
\bea
\delta \lambda^a &=& -\frac{i}{2}
\Sigma^{\acute{m}\acute{n}}\xi F_{\acute{m}\acute{n}}
=-\frac{i}{2}\{-\frac{i}{2}\ \mbox{\Large 1}\bigotimes
[\tilde{\Gamma}_{\acute{m}},\tilde{\Gamma}_{\acute{n}}]
\}\xi F^{a}_{\acute{m}\acute{n}}
\nonumber\\
 &=&-\frac{i}{2}F^{a}_{\acute{m}\acute{n}}\ \mbox{\Large 1} \bigotimes 
\ba{cc}
\bmat{cccc}
\tilde{\Sigma}^1_{\acute{m}\acute{n}}&&&\\
&\tilde{\Sigma}^2_{\acute{m}\acute{n}}&&\\
&&\tilde{\Sigma}^3_{\acute{m}\acute{n}}&\\
&&&\tilde{\Sigma}^4_{\acute{m}\acute{n}}\\
\emat 
&\bmat{c}
\xi_{\alpha i}\\
0\\
0\\
\bar{\xi}^{\dot{\alpha}i}\\
\emat
\ea
\nonumber\\
&=&-\frac{i}{2}F^{a}_{\acute{m}\acute{n}}
\bmat{c}
(\tilde{\Sigma}^1_{\acute{m}\acute{n}}\xi_{\alpha 1},
\tilde{\Sigma}^2_{\acute{m}\acute{n}}\xi_{\alpha 2},
\tilde{\Sigma}^3_{\acute{m}\acute{n}}\xi_{\alpha 3},
\tilde{\Sigma}^4_{\acute{m}\acute{n}}\xi_{\alpha 4})^T\\
0\\0\\
(\tilde{\Sigma}^1_{\acute{m}\acute{n}} \bar{\xi}^{\dot{\alpha}1},
\tilde{\Sigma}^2_{\acute{m}\acute{n}} \bar{\xi}^{\dot{\alpha}2},
\tilde{\Sigma}^3_{\acute{m}\acute{n}} \bar{\xi}^{\dot{\alpha}3},
\tilde{\Sigma}^4_{\acute{m}\acute{n}} \bar{\xi}^{\dot{\alpha}4})^T\\
\emat. 
\label{eqn:d10}
\eea
On the other hand using the self dual equation (\ref{eqn:d12})
\bea
\delta \lambda^a &=& -\frac{i}{2}
\Sigma^{\acute{m}\acute{n}}\xi \tilde{F}_{\acute{m}\acute{n}}
\nonumber\\
&=&-\frac{i}{2}F^{a}_{\acute{m}\acute{n}}\ \mbox{\Large 1} \bigotimes 
\ba{cc}
\bmat{cccc}
-\tilde{\Sigma}^1_{\acute{m}\acute{n}}&&&\\
&\tilde{\Sigma}^2_{\acute{m}\acute{n}}&&\\
&&-\tilde{\Sigma}^3_{\acute{m}\acute{n}}&\\
&&&\tilde{\Sigma}^4_{\acute{m}\acute{n}}\\
\emat 
&\bmat{c}
\xi_{\alpha i}\\
0\\
0\\
\bar{\xi}^{\dot{\alpha}i}\\
\emat
\ea
\nonumber\\
&=&-\frac{i}{2}F^{a}_{\acute{m}\acute{n}}
\bmat{c}
(-\tilde{\Sigma}^1_{\acute{m}\acute{n}}\xi_{\alpha 1},
\tilde{\Sigma}^2_{\acute{m}\acute{n}}\xi_{\alpha 2},
-\tilde{\Sigma}^3_{\acute{m}\acute{n}}\xi_{\alpha 3},
\tilde{\Sigma}^4_{\acute{m}\acute{n}}\xi_{\alpha 4})^T\\
0\\0\\
(-\tilde{\Sigma}^1_{\acute{m}\acute{n}} \bar{\xi}^{\dot{\alpha}1},
\tilde{\Sigma}^2_{\acute{m}\acute{n}} \bar{\xi}^{\dot{\alpha}2},
-\tilde{\Sigma}^3_{\acute{m}\acute{n}} \bar{\xi}^{\dot{\alpha}3},
\tilde{\Sigma}^4_{\acute{m}\acute{n}} \bar{\xi}^{\dot{\alpha}4})^T\\
\emat. 
\label{eqn:d10-1}
\eea
Thus the
$(\xi_{\alpha 1},\bar{\xi}^{\dot{\alpha}1})$ and 
$(\xi_{\alpha 3},\bar{\xi}^{\dot{\alpha}3})$ 
supersymmetry transformations leave the topological background
invariant while the
$(\xi_{\alpha 2},\bar{\xi}^{\dot{\alpha}2})$ and 
$(\xi_{\alpha 4},\bar{\xi}^{\dot{\alpha}4})$ do not.
We easily notice when anti-self dual equation is satisfied instead of 
(\ref{eqn:d12}) the unbroken supersymmetries and the broken ones 
exchange.

We have shown that there is a
six-dimensional object so called super 5-brane which breaks half 
of the supersymmetries in the background (\ref{eqn:d1})$\sim$
(\ref{eqn:d13}). 
Finally we would like to obtain the effective action for the 5-brane
using the method of nonliner realization.
In the low energty theory, 
there are four Nambu-Goldstone bosons of the broken translatons
and a $D=6$ Nambu-Goldstone Weyl spinor of the breaking supersymmetry.
The Nambu-Goldstone bosons of $SU(2)$ gauge symmetry 
are absorbed by the $D=6$ gauge fields 
and the $D=6$ gauge fields are massive.
Fortunately we know that the action for the Nambu-Goldstone fields 
is just the generalized Green-Schwarz
covariant superstring one \cite{HLP}\cite{AETW}\cite{BSTT}\cite{HM}:
\be
S=S_1+S_2
\label{eqn:ea1}
\ee
where $S_1$ is a supersymmetric Nambu-Goto action, and $S_2$ is a Wess-Zumino
action.
\be
S_1=\int d^6\sigma(-\sqrt{ -h})
\label{eqn:ea2}
\ee
where $h=\det h_{mn},\ h_{mn}=\Pi_{m}^{M}\Pi_{nM},
\ \Pi_{m}^{M}=\partial_m x^M-i\bar{\theta}\Gamma^M\partial_m\theta$.
Here $m,n=1,...6$ are world sheet indices and $M=1,...10$ is
a spacetime index. The Fermi field $\theta$ is a scalar on the world
sheet and a Majorana-Weyl spinor in $D=10$.
\be
S_2=\int_{WS}\Omega_6
\label{eqn:ea3}
\ee
where $\Omega_6$ is given by $\Omega_7$ as $d\Omega_6=\Omega_7$.
 The closed supersymmetric seven-form $\Omega_7$ is unique up to
 normalization and given by
\be
\Omega_7=\Pi^I\Pi^J\Pi^K\Pi^L\Pi^M d\bar{\theta}\Gamma_{IJKLM}d\theta.
\label{eqn:ea4}
\ee
The action (\ref{eqn:ea1}) is invariant under the nonlinear supersymmetry 
\be
\delta\theta = \xi,\quad \delta x^M=i\bar{\xi}\Gamma_M\theta.
\label{eqn:ea5}
\ee
However there is a local fermionic symmetry, this symmetry involves
half of the nonlinear action of $\xi$, so only half of the
supersymmetries (\ref{eqn:ea5}) act $\theta$ nonlinearly.
Similarly all the translations act nonlinearly for Bose fields $x^M$. 
However six of the translations are equivarent to
coordinate translations on the membrane ground state. 
As the same way, we shall be able to get $D=9\ $ 4-brane, 
$D=8\ $ 3-brane,$\cdots$ and $D=6\ $ 1-brane.\cite{AETW}


\vspace{2cm}
\noindent{\bf Acknowledgement}

We would like to thank S. Wada for useful discussions, especially 
we would like to thank
K. Sugiyama and T. Uematsu for numerous useful discussions.



\newpage
\baselineskip 17pt
\begin{thebibliography}{99}

\bibitem{HP1}
     J.~Hughes and J.~Polchinski, {\sl Nucl.~Phys.} {\bf B278} (1986)
     147.

\bibitem{NO}
     H.~B.~Nielsen and P.~Olesen, {\sl Nucl.~Phys.} {\bf B61} (1973) 45.

\bibitem{GS}
     M.~B.~Green and J.~H.~Schwarz {\sl Phys.~Lett.} 
     {\bf 136B} (1984) 367.

\bibitem{HLP}
     J.~Hughes, J.~Liu and J.~Polchinski, {\sl Phys.~Lett.} 
     {\bf 180B} (1986) 370.

\bibitem{BSS}
     L.~Brinkand, J.~H.~Schwarz and J.~Schrk,  {\sl Nucl.~Phys.} 
     {\bf B121} (1977) 77.

\bibitem{AETW}
     A.~Ach{\'u}carro, J.~M.~Evans, P.~K.~Townsend and D.~L.~Wiltshire,\\ 
     {\sl Phys.~Lett.} {\bf 198B} (1987) 441.

\bibitem{BSTT}
     E.~Berghoeff, E.~Sezgin, Y.~Tanii and P.~K.~Townsend,\\ 
     {\sl Ann.~Phys.} {\bf 199} (1990) 340.
   
\bibitem{SOH}
     M.~Sohnius, {\sl Phys.~Rep.} {\bf 128} (1985) 39.

\bibitem{STR}
     J.~Strathdee, {\sl Int.~J.~Mod.~Phys.} {\bf A,~Vol.~2,~No.~1} (1987) 273.

\bibitem{HM}
     M.~Henneaux and L.~Mezincescu,  {\sl Phys.~Lett.} 
     {\bf 152B} (1985) 340.

\end{thebibliography}
\end{document}